\documentclass[conference]{IEEEtran}
\IEEEoverridecommandlockouts
\usepackage{hyperref}
\usepackage{cite}
\usepackage{bm}
\usepackage{amsmath,amssymb,amsfonts}
\usepackage{algorithmic}
\usepackage{graphicx}
\usepackage{textcomp}
\usepackage{xcolor}

\usepackage{tikz}

\usepackage{tikz}

\usetikzlibrary{shapes.geometric}
\usetikzlibrary{shapes.misc}
\usetikzlibrary{shapes.symbols}
\usetikzlibrary{arrows, arrows.meta, positioning, 3d,calc}
\usetikzlibrary[math, decorations.pathreplacing]
\tikzset{
  >=stealth',
}

\usepackage[straight voltages, american]{circuitikz}
\ctikzset{resistors/scale=0.7,
  capacitors/scale=0.7,
  inductors/scale=0.7,
  instruments/scale=0.5,
}

\makeatletter
\tikzset{self loop/.style =  {to path={
  \pgfextra{}
  [looseness=12,min distance=10mm]
  \tikz@to@curve@path},font=\sffamily\small
  }}
\makeatother

\makeatletter
\tikzset{%
  block/.style      = {draw, thick, rectangle, minimum height = .5cm, minimum width = 1cm, align=center, font=\footnotesize},
  hub/.style        = {draw, circle, fill=black, minimum size=.10cm, inner sep=0.00cm},
  symhub/.style     = {hub, fill=none, draw, inner sep=0.005cm, scale=.5},
  connection/.style = {draw, ->},
  conlabel/.style   = {scale=0.75},
  ede/.style = {circle, fill=lightgray, font=\scriptsize, inner sep=0.05cm, minimum size=.25cm},
  eae/.style = {draw, circle, font=\scriptsize, inner sep=0.05cm, minimum size=.25cm},
}
\makeatother

\makeatletter
\newcommand{\pgfwest}{%
    \pgf@process{\northeast}%
    \pgf@ya=.5\pgf@y%
    \pgf@process{\southwest}%
    \pgf@y=.5\pgf@y%
    \advance\pgf@y by \pgf@ya%
}
\newcommand{\pgfeast}{%
    \pgf@process{\southwest}%
    \pgf@ya=.5\pgf@y%
    \pgf@process{\northeast}%
    \pgf@y=.5\pgf@y%
    \advance\pgf@y by \pgf@ya%
}
\newcommand{\pgfsouth}{%
    \pgf@process{\northeast}%
    \pgf@xa=.5\pgf@x%
    \pgf@process{\southwest}%
    \pgf@x=.5\pgf@x%
    \advance\pgf@x by \pgf@xa%
}
\newcommand{\pgfnorth}{%
    \pgf@process{\southwest}%
    \pgf@xa=.5\pgf@x%
    \pgf@process{\northeast}%
    \pgf@x=.5\pgf@x%
    \advance\pgf@x by \pgf@xa%
}

\long\def\pgfshapeaddanchor#1#2{%
{%
  \def\pgf@sm@shape@name{#1}%
  \let\anchor=\pgf@sh@anchor%
  #2}%
}
\pgfshapeaddanchor{rectangle}{%
  \anchor{west north west}{%
    \pgf@process{\northeast}%
    \pgf@ya=1.5\pgf@y%
    \pgf@process{\southwest}%
    \pgf@y.5\pgf@y%
    \advance\pgf@y by \pgf@ya%
    \pgf@y=.5\pgf@y%
  }%
  \anchor{north north west}{%
    \pgf@process{\southwest}%
    \pgf@xa=1.5\pgf@x%
    \pgf@process{\northeast}%
    \pgf@x.5\pgf@x%
    \advance\pgf@x by \pgf@xa%
    \pgf@x=.5\pgf@x%
  }%
  \anchor{west south  west}{%
    \pgf@process{\northeast}%
    \pgf@ya=.5\pgf@y%
    \pgf@process{\southwest}%
    \pgf@y=1.5\pgf@y%
    \advance\pgf@y by \pgf@ya%
    \pgf@y=.5\pgf@y%
  }%
  \anchor{south south  west}{%
    \pgf@process{\northeast}%
    \pgf@xa=.5\pgf@x%
    \pgf@process{\southwest}%
    \pgf@x=1.5\pgf@x%
    \advance\pgf@x by \pgf@xa%
    \pgf@x=.5\pgf@x%
  }%
  \anchor{west -1}{%
    \pgfwest%
    \advance\pgf@y by -0.5 cm%
  }%
  \anchor{west 1}{%
    \pgfwest%
    \advance\pgf@y by 0.5 cm%
  }%
  \anchor{west A}{%
    \pgfwest%
    \advance\pgf@y by 0.25 cm%
  }
  \anchor{west B}{%
    \pgfwest%
    \advance\pgf@y by -0.25 cm%
  }%
  \anchor{east -1}{%
    \pgfeast%
    \advance\pgf@y by -0.5 cm%
  }%
  \anchor{east 1}{%
    \pgfeast%
    \advance\pgf@y by 0.5 cm%
  }%
  \anchor{east A}{%
    \pgfeast%
    \advance\pgf@y by 0.25 cm%
  }
  \anchor{east B}{%
    \pgfeast%
    \advance\pgf@y by -0.25 cm%
  }%
  \anchor{north -1}{%
    \pgfnorth%
    \advance\pgf@x by -0.5 cm%
  }%
  \anchor{north 1}{%
    \pgfnorth%
    \advance\pgf@x by 0.5 cm%
  }%
  \anchor{north A}{%
    \pgfnorth%
    \advance\pgf@x by -0.25 cm%
  }
  \anchor{north B}{%
    \pgfnorth%
    \advance\pgf@x by 0.25 cm%
  }%
  \anchor{south -1}{%
    \pgfsouth%
    \advance\pgf@x by -0.5 cm%
  }%
  \anchor{south 1}{%
    \pgfsouth%
    \advance\pgf@x by 0.5 cm%
  }%
  \anchor{south A}{%
    \pgfsouth%
    \advance\pgf@x by 0.125 cm%
  }
  \anchor{south B}{%
    \pgfsouth%
    \advance\pgf@x by -0.125 cm%
  }%
}
\makeatother

\def\BibTeX{{\rm B\kern-.05em{\sc i\kern-.025em b}\kern-.08em
    T\kern-.1667em\lower.7ex\hbox{E}\kern-.125emX}}
\begin{document}

\newcommand{\mrm}[1]{\mathrm{#1}}

\newcommand{\dq}{{dq}}
\newcommand{\ifilt}{i_\mrm{f}}
\newcommand{\iout}{i_\mrm{g}}
\newcommand{\uout}{v_\mrm{C}}

\newcommand{\Pref}{P^\mrm{s}}
\newcommand{\Qref}{Q^\mrm{s}}

\newcommand{\Vref}{V^\mrm{s}}
\newcommand{\wref}{\omega^\mrm{s}}
\newcommand{\cf}{\eta}

\newcommand{\cfpred}{\dot{\Theta}^\mrm{pred}}
\newcommand{\cfmeas}{\dot{\Theta}^\mrm{m}}

\newcommand{\cp}{\Theta}
\newcommand{\cppred}{\Theta^\mrm{pred}}

\newcommand{\xc}{x_\mrm{c}}

\newcommand{\thetabar}{{\overline{\theta}}}
\newcommand{\Thetabar}{{\overline{\Theta}}}

\newcommand{\ibar}{{\overline{\imath}}}
\newcommand{\vbar}{{\overline{v}}}
\newcommand{\Sbar}{{\overline{S}}}

\newcommand{\sbar}{{\overline{s}}}
\newcommand{\Cbar}{{\overline{C}}}

\newcommand{\etabar}{{\overline{\cf}}}
\newcommand{\zbar}{{\overline{z}}}
\newcommand{\Ybar}{\overline{Y}}

\newcommand{\bmat}[1]{\begin{bmatrix} #1 \end{bmatrix}}
\newcommand{\anna}[1]{\textcolor{green}{#1}}
\newcommand{\frank}[1]{\textcolor{dblue}{#1}}
\newcommand{\jakob}[1]{\textcolor{purple}{#1}}
\newcommand{\robin}[1]{\textcolor{orange}{[RD: #1]}}

\bibliographystyle{IEEEtran}

\title{Complex Phase Analysis of Power Grid Dynamics
\thanks{Federal Ministry for Economic Affairs and Climate Action: Projects OpPoDyn (03EI1071A) and CoCoMusy (360332943); and DFG: MATH+ (EXC-2046/1, project ID: 390685689)}
}

\author{
    \IEEEauthorblockN{
        Jakob Niehues\IEEEauthorrefmark{1}\IEEEauthorrefmark{2} \thanks{Corresponding author: jakob.niehues\{ät\}pik-potsdam.de} \thanks{ORCID: J.N. 0000-0003-0140-6910; A.B. 0000-0002-3555-8173; F.H. 0000-0001-5635-4949},
        Anna Büttner\IEEEauthorrefmark{1} ,
        Anne Riegler\IEEEauthorrefmark{1} and
        Frank Hellmann\IEEEauthorrefmark{1}
    }
    \IEEEauthorblockA{\IEEEauthorrefmark{1} Potsdam Institute for Climate Impact Research (PIK), Member of the Leibniz Association,\\ P.O. Box 60 12 03, D-14412, Potsdam, Germany}
    \IEEEauthorblockA{\IEEEauthorrefmark{2} Institut für Mathematik, Technische Universität Berlin, ER 3-2, Hardenbergstrasse 36a, 10623 Berlin, Germany}
}

\maketitle
\thispagestyle{plain}
\pagestyle{plain}

\begin{abstract}
    With an increasing share of renewable energy sources, accurate and efficient modeling of grid-forming inverters is becoming crucial for system stability.
    Linear methods are a powerful tool for understanding dynamics close to an operating point, but usually depend on the reference trajectory.
    Thus, small deviations can render linear models invalid over time, posing a significant challenge in practice, and complicating theoretical analysis.
    As a solution, we show that the complex phase offers a robust formulation independent of reference phases and frequencies, thus preserving invariance properties under linearization.
    This enables robust system identification during realistic conditions and opens the road to powerful stability analysis of inverter-based grids.
\end{abstract}


\section{Introduction}
Renewable energy sources (RESs) and the power-electronic inverters that connect them to the grid play an increasingly important role in the electric power mix. As power systems increasingly transition to RES-dominated configurations, grid-forming inverters have become a critical technological solution for maintaining grid stability without the presence of synchronous generation \cite{matevosyan_grid-forming_2019}. Given their significance in RES-dominated power grids, grid-forming inverters must be modeled appropriately, and their interactions and stability properties must be rigorously understood. 

Many powerful techniques for modeling and analyzing systems involve linearizing the dynamics of the voltage $v$ and current $\imath$ around an operating state $v^\circ$, $\imath^\circ$. However, this linearization results in linear-time-periodic (LTP) systems, which are mathematically complex compared to time-invariant systems. To address this, the system is typically transformed into a frame co-rotating at the operating state's frequency $\omega^\circ$, yielding a linear-time-invariant (LTI) system. While this approach simplifies analysis, it is not without limitations.

The co-rotating frame transformation requires a precise frequency measurement and assumes constant frequency $\omega^\circ$, an assumption which is often invalid in real-world scenarios. Furthermore, the transformation introduces a critical limitation: The LTI system depends on an arbitrary initial phase $\varphi^\circ$, which renders the linearization non-unique. Hence, such linearizations only apply if the phase stays near the linearization point $\varphi^\circ$, and frequency deviations, even if minimal, can invalidate the linear model. Consider a scenario during which a temporary power mismatch occurs: After restoring the power balance, the operating state is the same as the initial $v^\circ$ but shifted by a phase, necessitating a new linearized system. This is a problem for the system identification of real components. The dependence on an arbitrary phase also complicates the interpretation of linear stability results in terms of machine parameters.

In the following, we will show that these issues are solved by working in terms of complex frequency~\cite{milano_complex_2022} and complex phase~\cite{buttner_complex-phase_2024}.
The concept of complex frequency, and the derived concept of complex phase that we will leverage in the following, was originally introduced by Milano \cite{milano_complex_2022} and has been explored extensively lately. It simplifies the equations for power flow \cite{milano_complex_2022}, has been employed for analyzing and improving control \cite{moutevelis_design_2023,milano_enhancing_2024,bernal_improving_2024}, taxonomy of models \cite{moutevelis_taxonomy_2024}, and estimating inertia \cite{zhong_-line_2022}. It allows the analysis of complex oscillator-based grid-forming control concepts \cite{he_quantitative_2024,he_complex-frequency_2024,milano_dual_2025}, and recasting the dynamics of power grids as adaptive network equations well suited for bilinear control \cite{buttner_complex_2024}. Further, \cite{kogler_normal_2022} showed that under very broad assumptions, grid-forming devices can be modeled by specifying how the complex frequency reacts to deviations in active power, reactive power, and voltage amplitude.

By formulating linearized dynamics of devices with respect to complex phases and power variables, we obtain LTI systems describing the device dynamics that:
\begin{enumerate}
    \item Eliminate the need for co-rotating frame transformations,
    \item Remove dependency on arbitrary absolute phase,
    \item Provide a more robust modeling approach for grid-forming components.
\end{enumerate}
The relationship between voltage and current is linear, and linearizing the device in these variables automatically linearizes the full interconnected system. This is not true when we linearize devices using complex phases and power. As the relationship between power and complex phase is non-linear, the full system retains important physical non-linearities. 

The remainder of this paper is structured as follows: In Section \ref{sec:problem}, we review in detail how linearizing introduces an arbitrary phase and how phase shifts in non-linear and linear systems differ. Section \ref{sec:solution} reviews the notions of complex phase and complex frequency, and demonstrates that these variables, together with ideas from the normal form description of grid-forming actors \cite{kogler_normal_2022}, resolve these issues.

In Sections \ref{sec:data-driven identification} and \ref{sec:stability analysis}, we demonstrate the advantages of the complex phase description through recent system identification and linear stability results. The results for the system identification include an example of an inverter in a power-hardware in-the-loop laboratory with phase drifts. 
The linear stability results depend on general dynamical characteristics of the system that can be easily related to model parameters, and are independent of absolute phases.
Both results have been enabled by the unique suitability of complex phase variables for applying linear methods to power grid dynamics.

\section{Phase dependence and linearization}
\label{sec:problem}
To demonstrate how the complex phase approach differs from working directly in terms of voltages, this section will show how a power grid component can be linearized in terms of voltage and current. We consider a grid-forming inverter that acts as a voltage source $v$ that reacts to the incoming current $\imath$. We describe the system in $\alpha \beta \gamma$-coordinates and assume that the $\gamma$ coordinate $v_\gamma = 0$, meaning that the instantaneous sum of the three voltage phases is zero. For this section, we neglect the internal states of the grid-forming device entirely and assume that it acts as a voltage source with smooth changes in the voltage, such that we can write:
\begin{align}
    \dot v &:= \bmat{\dot v_\alpha\\ \dot v_\beta}
    = \bmat{f_\alpha(v, \imath) \\ f_\beta(v, \imath) } = f (v, \imath)\label{eq:voltage_definition} ,
\end{align}
with the dynamics given by a nonlinear function $f$.
The aim is to understand the behavior of $f$ near a reference trajectory around which we linearize. This is typically an operating state $v^\circ(t)$ of the system with frequency $\omega^\circ$. Such operating states can be written using the rotation matrix $R$:
\begin{align}
    R(\Phi) &:= \bmat{\cos\Phi & -\sin\Phi\\ \sin\Phi & \cos\Phi}
\end{align}
as
\begin{align}
    v^\circ(t) &:= R(\omega^\circ t) v^\circ(0) \; .\label{eq:stationary}
\end{align}

For small deviations $\delta v(t) := v(t) - v^\circ(t)$, we have:
\begin{align}
\label{eq:normal form v alpha beta linearized}
    \bmat{
        \delta\dot v_\alpha\\
        \delta\dot v_\beta}
        &= J^\circ \bmat{\delta v_\alpha \\ \delta v_\beta}
        + D^\circ \bmat{\delta\imath_\alpha \\ \delta \imath_\beta},
\end{align}
with the input matrix $D^\circ = D(v^\circ(t), \imath^\circ(t))$ and Jacobian $J^\circ=J(v^\circ(t), \imath^\circ(t))$ given in terms of:
\begin{align}\label{eq:normal form vi linearized J}
    J &:=
    \bmat{
    J_{\alpha\alpha} & J_{\alpha\beta} \\
    J_{\beta\alpha} & J_{\beta\beta} },
\end{align}
where $J_{\alpha\beta} :=  \frac{d f_\alpha}{d v_\beta}$ and analogously $D_{\alpha\beta}:=\frac{d f_\alpha}{d \imath_\beta}$ etc.

Equation \eqref{eq:normal form v alpha beta linearized} is a linear-time-periodic system, as the coefficients in $J^\circ$ and $D^\circ$ explicitly depend on time. The time dependence of the reference trajectory can be avoided by switching from the $\alpha \beta$-coordinates to a rotating $dq$-frame using the Park transformation \cite{machowski_power_2012}:
\begin{align}
    v^{dq} &:= \bmat{v_d \\ v_q} := R(- \Omega t) v\\
    v &= R^\top(- \Omega t) v^{dq} = R (\Omega t) v^{dq}.
\end{align}

If the transformation frequency $\Omega$ matches $\omega^\circ$, then $v^{\circ, dq} = v^\circ(0)$ is time independent. A direct calculation shows that $\frac{\partial}{\partial t} R(- \omega^\circ t) = - \omega^\circ R({\pi}/{2}) R(- \omega^\circ t)$. Using this and the definition of $v^{dq}$ and equation \eqref{eq:voltage_definition}, the time evolution of $v^{dq}$ is given by:
\begin{align}
    \dot v^{dq} &= \left(\frac{\partial}{\partial t} R(- \omega^\circ t)\right) v + R(- \omega^\circ t) f(v, \imath) \\
    &= - \omega^\circ R({\pi}/{2}) v^{dq} + R(- \omega^\circ t) f(R(\omega^\circ t) v^{dq}, R(\omega^\circ t) \imath^{dq}) \label{eq:vdq_dynamcs_phase}.
\end{align}

The right-hand side is independent of time for all $v^{dq}$ and $\imath^{dq}$ if and only if the nonlinear function $f$ defining the voltage dynamics fulfills the following condition:
\begin{align}
R(- \omega^\circ t) f(R(\omega^\circ t) v^{dq}, R(\omega^\circ t) \imath^{dq}) = f(v^{dq}, \imath^{dq})
\end{align}
for all times $t$, and thus all angles $\Phi$:
\begin{align}
R(- \Phi) f(R(\Phi) v^{dq}, R(\Phi) \imath^{dq}) = f( v^{dq}, \imath^{dq})\label{eq:phase shift invariance}.
\end{align}
Using this condition in equation \eqref{eq:vdq_dynamcs_phase} we obtain:
\begin{align}
    \dot v^{dq} = f^{dq}(v^{dq}, \imath^{dq}) := - \omega^\circ R({\pi}/{2}) v^{dq} + f(v^{dq}, \imath^{dq}) 
    \label{eq:nonlinear dq}.
\end{align}

Due to the phase condition \eqref{eq:phase shift invariance}, the phase with respect to which we describe the system is arbitrary. This can be observed by considering the $dq$-system rotated by a constant phase ${v'}^{dq}=R(\Phi) v^{dq}$. Due to the phase condition, we can cancel the $R(\Phi)$ in the equation for $\dot {v'}^{dq}$ everywhere, and ${v'}^{dq}$ and $v^{dq}$ have the same equations: All phase choices $\Phi$ lead to the same dynamics $f^{dq}$.

This is no longer true for the linearized system. Linearizing equation \eqref{eq:nonlinear dq} gives the following system:
\begin{align}
\label{eq:normal form v dq linearized}
\delta \dot v^{dq} &= J^{\circ, dq} \delta v^{dq} + D^{\circ, dq} \delta \imath^{dq}\\
J^{\circ, dq} &= J(v^{\circ, dq}, \imath^{\circ, dq}) - \omega^\circ R({\pi}/{2}) \\
D^{\circ, dq} &= D(v^{\circ, dq}, \imath^{\circ, dq}).
\end{align}
from which we can see that $J^{\circ, dq}$ and $D^{\circ, dq}$ explicitly depend on the linearization point ${v^{\circ, dq}}$, and thus change with $\Phi$. Hence, linearization introduces a dependence on an absolute reference phase of the linearization point and creates a continuum of linear systems related by phase shifts.

However, there is a linearized version of the phase shift under which the linearized system is invariant. Consider a small rotation $v' = R(\epsilon) v^{\circ, dq}$. We use \eqref{eq:phase shift invariance} and $R(\Phi) R(-\Phi) = I$ the identity matrix to write:
\begin{align}
    f^{dq}(R(\epsilon) v^{\circ, dq}, R(\epsilon) \imath^{\circ, dq}) = R(\epsilon) f^{dq}(v^{\circ, dq}, \imath^{\circ, dq}).
\end{align}

As the reference trajectory is a solution of the dynamical equations, we have $f^{dq}(v^{\circ, dq}, \imath^{\circ, dq}) = 0$, and the right hand side is $0$. Expanding the left hand side in $\epsilon$, using $R(\epsilon) = 1 + \epsilon R(\pi/2)$, one can check that:
\begin{align}
    0 = \epsilon J^{\circ, dq} R(\pi/2) v^{\circ, dq} + \epsilon D^{\circ, dq} R(\pi/2) \imath^{\circ, dq}.
\end{align}

The vectors $R(\pi/2) v^{\circ, dq}$ and $R(\pi/2) \imath^{\circ, dq}$ point in the direction of the phase shift at the linearization point $v^{\circ, dq}$. The linear equations \eqref{eq:normal form v alpha beta linearized} and \eqref{eq:normal form v dq linearized} are unaffected by shifting $\delta v^{dq}$ and $\delta \imath^{dq}$ any distance in this direction.

\begin{figure}[htbp]
    \centerline{\includegraphics[width=\columnwidth]{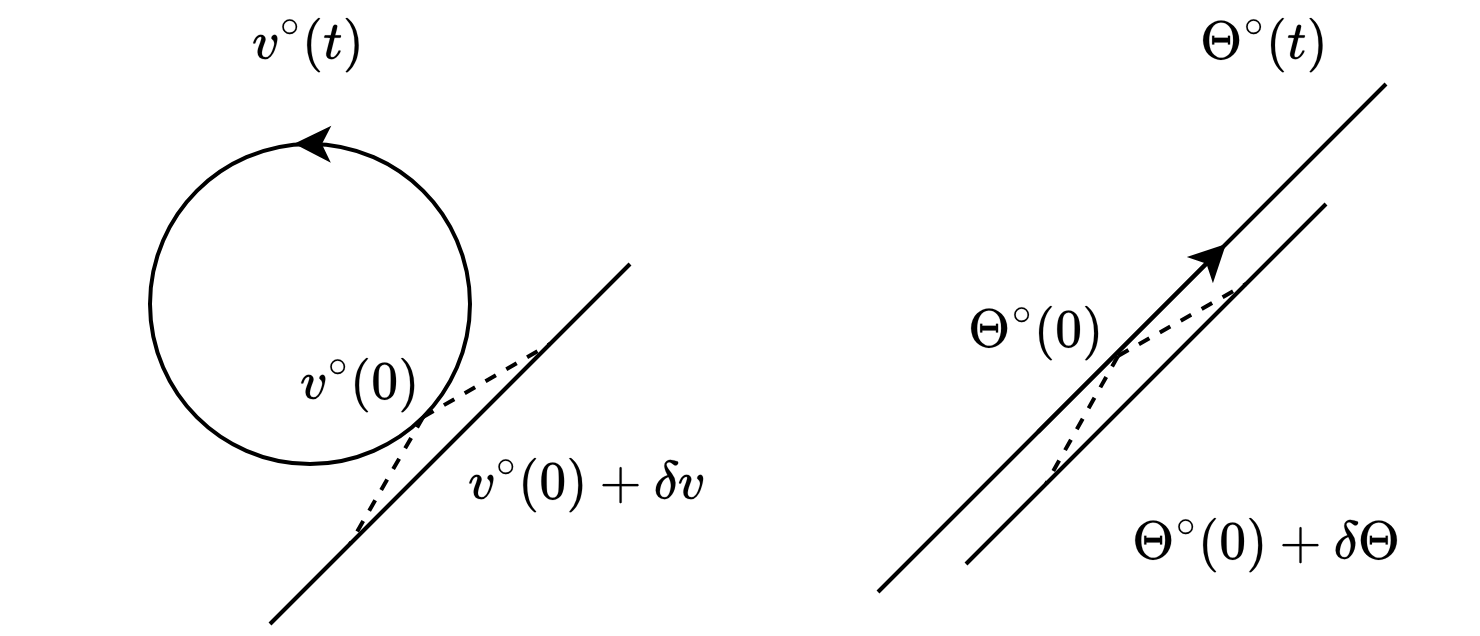}}
    \caption{Linearization of a phase shift of the voltage at a given reference voltage $v^\circ(0)$ on the steady state trajectory $v^\circ(t)$, and the phase shift of the complex phase at $\Theta^\circ(0)$ on the steady state trajectory $\Theta^\circ(t)$. The linearized phase shifts are slightly offset from the linearization point for clarity (dashed lines). For the complex phase (right), the steady state trajectory and the linearized phase shift coincide, both leave the linearization and the full model invariant. For the voltage (left), the linearized phase shift diverges from the steady state trajectory and the non-linear phase shift. The linearization is invariant under linearized phase shifts but not under non-linear phase shifts. Different linearization points on the same circle give different linearized dynamics.}
\label{fig:coordinates and linearization}
\end{figure}

This mismatch between the phase shifts of the non-linear and the linear equations is at the core of the problems outlined above. The invariant shift in the linear system does not correspond to the invariant shift of the full system, except at the linearization point, as illustrated on the left side of Fig.~\ref{fig:coordinates and linearization}. This poses a serious challenge for many practical applications. If there is a slight mismatch between the $dq$-frame frequency and $\omega^\circ$, for example, due to a small power mismatch, the $dq$-voltage will drift away from the linearization point. The shifts observed in the linearized system will increasingly diverge from those in the non-linear one, and the linearized equations will lose their validity. They will remain invalid even after the power mismatch is resolved.

\section{Complex Phase Linearization}
\label{sec:solution}
In this section, we show how the complex phase resolves the issues of differently behaved nonlinear and linear dynamics, and results in an LTI system independent of the reference phase at the linearization point. Throughout, complex quantities will be written with a hat: $\hat z$. We denote the imaginary unit $j$ and the complex conjugate $\hat z^*$.

We begin by recalling the complex phase description of the voltage \cite{buttner_complex_2024}.
Any vector $v$ has an instantaneous phase and amplitude. A convenient way to work in terms of these is to translate the vector $v$ into a complex number
\begin{align}
    \hat v &:= v_\alpha + j v_\beta \, ,
\end{align}
which allows us to write the phase in terms of the complex logarithm $\ln$:
\begin{align}
    V &:= |\hat v|\\
    \varphi &:= \arg(\hat v) = \Im(\ln \hat v(t)).
\end{align}
If the voltage changes smoothly, we can also define an instantaneous notion of frequency in terms of the velocity of $\hat v(t)$:
\begin{align}
    \omega(t) := \frac{d}{d t} \Im(\ln \hat v(t)) = \Im\frac{\dot {\hat v}}{\hat v}\;.
\end{align}

In \cite{milano_complex_2022}, Milano suggests extending this notion of instantaneous frequency to the complex frequency with $\rho := \dot V/V$:
\begin{align}
    \hat \eta(t) &:= \rho(t) + j \omega(t) := \frac{d}{d t} \ln(\hat v(t)) = \frac{\dot {\hat v}}{\hat v}\\
    \eta &:= [\rho, \omega]^\top.
\end{align}
This motivates us to introduce the complex phase:
\begin{align}
\hat \Theta &:= \sigma(t) + j \varphi(t) := \ln(\hat v(t)) \\
\Theta &:= [\sigma, \varphi]^\top,
\end{align}
such that $\dot \Theta = \eta$.
The operating states of the form $v^\circ = R(\omega^\circ t + \varphi^\circ) [e^{\sigma^\circ}, 0]^\top$ have complex phase
\begin{align}
{\hat \Theta}^\circ = \sigma^\circ + j ( \omega^\circ t + \varphi^\circ).
\end{align}
The rotating motion of the operating state in the $\alpha \beta$-plane is mapped to a linear motion in the complex phase plane as we can also see in Fig.~\ref{fig:coordinates and linearization}. As we will see, this allows us to obtain one linearized system for all $\omega^\circ$, $\varphi^\circ$.

Instead of working in terms of $f(v, \imath)$ as above, we now work equivalently in terms of complex phase, complex frequency, and the active power $P$ and reactive power $Q$. The dynamics are determined by the complex frequency's dependence on the other variables:
\begin{align}
    \dot {\hat v} = \hat v \hat \eta(\sigma, \varphi, P, Q),
\end{align}
and as $\hat v = \exp(\hat \Theta)$ and $\hat \imath^* = (P + j Q)/\hat v$, this is just a coordinate transformation which contains the same information as \eqref{eq:voltage_definition}. The reason for working in terms of $P$ and $Q$ rather than in terms of a direct variation of the current is that $P$ and $Q$ are invariant under phase shifts. Using this, it can then be checked that the phase behavior of $f$ (see \eqref{eq:phase shift invariance}) implies
\begin{align}
\eta(\sigma, \varphi, P, Q) = \eta(\sigma, \varphi + \Phi, P, Q) = \eta(\sigma, P, Q).
\end{align}
That is, the complex frequency is independent of the imaginary part of the complex phase. Let us now observe what happens if we linearize the system around $\Theta^\circ$: 
\begin{align}\label{eq:linearized complex frequency dynamics invariance}
    \bmat{\delta\dot\sigma\\ \delta\dot \varphi} = \underbrace{\bmat{J^\circ_{\sigma \sigma} & 0\\ J^\circ_{\varphi \sigma } & 0}}_{J^\eta} \bmat{\delta \sigma \\ \delta \varphi} + \underbrace{\bmat{D^\circ_{\sigma Q} & D^\circ_{\sigma P} \\ D^\circ_{\varphi Q} & D^\circ_{\varphi P}} }_{D^\eta} \bmat{\delta Q \\ \delta P}.
\end{align}

As $\eta$ is independent of $\varphi$, so are $J^\eta$ and $D^\eta$. And as the time dependence of the reference trajectory is entirely contained in $\varphi$, the above linear system is time invariant without any assumptions on $\omega^\circ$. In particular, $\omega^\circ$ can vary with time, but the linearization remains unchanged. The coefficients will depend on $P^\circ$, $Q^\circ$ and $\sigma^\circ = \ln(V^\circ)$ but not on any arbitrary phase. Thus, they directly represent the behavior of the device at a particular power flow.

In these coordinates, a phase shift leaves both the linearized and the full dynamics invariant. Further, the motion of the operating state and the direction of phase drifts (see right side of Fig.~\ref{fig:coordinates and linearization}) are along this invariant direction. In the scenario considered above, a small temporary power imbalance in the system will shift the linear and non-linear system in the same way. After it is resolved, the linear system will be as valid as it was prior to the phase shift.

Further, while we have linearized $\eta$ in terms of $P$, $Q$ and $\sigma$, there is no need a priori to linearize $P$ and $Q$ as a function of the $\Theta = \sigma + j \varphi$. If we keep the nonlinear dependency of $P$ and $Q$ on $\Theta$, the linearization \eqref{eq:linearized complex frequency dynamics invariance} corresponds to a subclass of the normal form approach introduced in \cite{kogler_normal_2022}.
This gives a more realistic description of the dynamics, and common control strategies, such as $P$-$\omega$ droop, can be represented exactly.

In the next two chapters, we highlight two applications of complex phase representation.

\section{System identification}
    \label{sec:data-driven identification}
    In \cite{buttner_complex-phase_2024}, a novel system identification approach is presented to derive low-dimensional models of grid-forming inverters. This approach builds on the complex phase and employs a Hammerstein-Wiener parametrization of the normal form model, introduced in \cite{kogler_normal_2022}. We will see below that this can be seen as a generalization and variation of \eqref{eq:linearized complex frequency dynamics invariance}.

    System identification refers to the process of constructing mathematical models of dynamical systems based on measured input-output data. A conventional choice for grid-forming inverters would be to consider the current $\imath$ and voltage $v$ as input and output variables, respectively. However, the same problems stated in section \ref{sec:problem} also introduce significant challenges here. The complex phase $\cp$ is used as the output variable to address these.

    \subsection{Hammerstein-Wiener Normal Form} 
    \label{sec:nf}
    In the previous section, the voltage dynamics introduced in equation \eqref{eq:voltage_definition} were derived under the assumption that the device has no internal degrees of freedom. This implies that the dynamics depend solely on the voltage or the corresponding complex phase. However, grid-forming inverters inherently incorporate energy storage elements, such as capacitors in a DC link, and utilize control strategies like droop control \cite{schiffer_conditions_2014}. These features introduce dynamic behavior that necessitates the inclusion of internal state variables, denoted as $\xc$, in the model. The normal form framework proposed in \cite{kogler_normal_2022} extends the formulation in equation \eqref{eq:voltage_definition} by explicitly accounting for these internal states.

\begin{figure}[htbp]
        \centering
        \begin{tikzpicture}
  \node[block, minimum width=0cm](he){$h^e$};
  \node[block, right=0.75cm of he, minimum height=.75cm, minimum width=0cm](LTI){LTI};
  \node[block, right=0.75cm of LTI, minimum width=0cm](int){$\int\mrm{d}t$};
  \node[block, right=0.75cm of int, minimum width=0cm, minimum height=.75cm,](exp){$\exp(\cdot)$};
  \node[block, right=0.75cm of exp, minimum height=.75cm, minimum width=0cm](PG){\shortstack{\scriptsize power\\\scriptsize grid}};

  \draw[connection](LTI.east) -- (int.west) node[pos=.5, above, conlabel]{$\cf$};
  \draw[connection](int.east) -- (exp.west) node[pos=.5, above, conlabel]{$\cp$};
  \draw[connection](exp.east) -- (PG.west) node[pos=.5, above, conlabel]{$v$};

  \draw[connection](he) -- (LTI) node[conlabel, pos=.5, above]{$e$};
  \draw[connection](-1.75,0) -- (he) node[conlabel, pos=.5, below]{$\Pref, \Qref, \Vref$};
  \draw[connection](PG.east)--++(0.25,0) --++(0,-1.25) -| (he.south B) node[conlabel, pos=0.25,above](){$i$};
  \draw[connection](exp.east)--++(0.375,0) node[hub]{} --++(0,-0.75) -| (he.south A) node[conlabel, pos=0.25,above](){$v$};

\path (he.west) --++ (-0.125,0.5) coordinate (cb1a); 
\path (he.east) --++ ( 0.125,0.5) coordinate (cb1b); 
\draw[decorate, decoration={brace, amplitude=1ex, raise=0ex}] (cb1a) -- (cb1b) node[pos=.5, above=1ex, text width=1.75cm, scale=0.75, align=center] {input\\ nonlinearity};

\path (LTI.west) --++ (-0.125,0.5) coordinate (cb1a); 
\path (int.east) --++ ( 0.125,0.5) coordinate (cb1b); 
\draw[decorate, decoration={brace, amplitude=1ex, raise=0ex}] (cb1a) -- (cb1b) node[pos=.5, above=1ex, text width=1.75cm, scale=0.75, align=center] {linear\\ subsystem};

\path (exp.west) --++ (-0.125,0.5) coordinate (cb1a); 
\path (exp.east) --++ ( 0.125,0.5) coordinate (cb1b); 
\draw[decorate, decoration={brace, amplitude=1ex, raise=0ex}] (cb1a) -- (cb1b) node[pos=.5, above=1ex, text width=1.75cm, scale=0.75, align=center] {output\\ nonlinearity};
\end{tikzpicture}
        \caption{A single grid-forming inverter, modeled by the normal form, coupled to the power grid.}
        \label{fig:normal form}
    \end{figure}
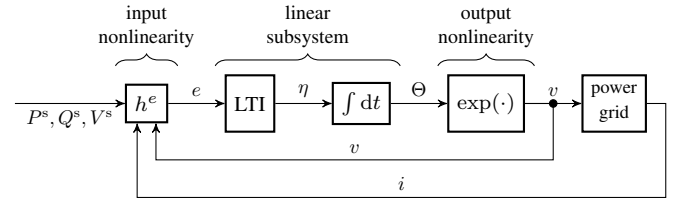

    To introduce the normal form, the quantities $P$, $Q$, and $V$ are considered with their respective error coordinates $e$ relative to the set-points $\Pref$, $\Qref$, and $\Vref$:
\begin{align}
    \begin{aligned} 
        e &:= h^e\left(v, \imath, \Pref, \Qref, \Vref\right)\,   \\
        &:= [P - \Pref, Q - \Qref, V - \Vref]^\top.
        \label{eq:err}
    \end{aligned}
\end{align}
    This yields the following most universal form of the normal form model:
    \begin{align}
    \label{eq:normal form system identification x}
        \dot{x}_\mrm{c}(t) &= g(e, \xc)\,,\\
        \label{eq:normal form system identification  eta}
       \cf(t) &= f(e, \xc)\,, \\
       \label{eq:normal form system identification theta}
       \dot{\cp}(t) &= \cf \,, \\
       \label{eq:normal form system identification output}
       \hat v(t) &= e^{\hat \cp(t)} \,, 
    \end{align}
    where $f$ and $g$ are continuous, non-linear functions. If $f$ and $g$ are linear, the normal form exhibits a Hammerstein-Wiener (H-W) structure. H-W models are characterized by a static non-linearity at the input, followed by a linear subsystem that defines the system dynamics, and ending with a non-linearity that computes the output \cite{ljung_system_1998}. In the normal form, the input non-linearity is given by equation \eqref{eq:err}. The output non-linearity is defined by the computation of the $dq$-voltage via the complex phase \eqref{eq:normal form system identification output}. The H-W normal form is defined as:
    \begin{align}
        \label{eq:normal form Hammerstein Wiener x}
        \dot{x}_\mrm{c} &= A \xc + B e\,, \\
        \label{eq:normal form Hammerstein Wiener eta}
        \cf &= C \xc + D e\, .
    \end{align}
    The structure of the H-W normal form is summarized in Fig.~\ref{fig:normal form}. The H-W model structure is commonly employed for system identification approaches \cite{wills_identification_2013} as non-linearities are maintained, but parameter estimation is simplified \cite{ljung_system_1998}. 
    Any structural differences between grid-forming inverters are encapsulated in the parameter matrices $A$, $B$, $C$, and $D$, which can be identified from measured data. 

    \subsection{Results}
\begin{figure}[htbp]
      \centering
      \includegraphics[trim={0 5 0 35}, clip, width=3.2in]{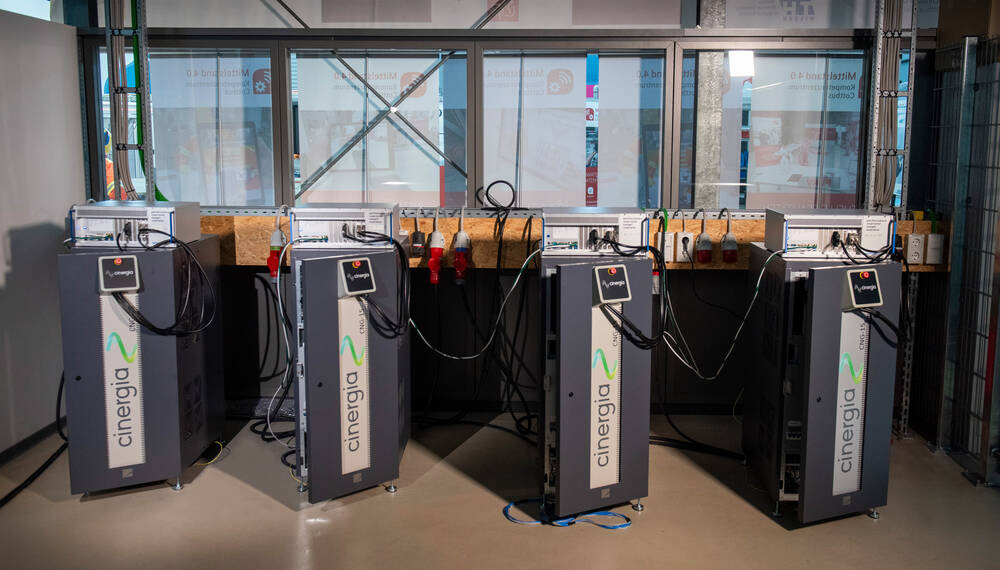}
      \caption{Four inverters in the power hardware in the loop lab.}\label{fig:labphoto}
    \end{figure}

    In \cite{buttner_complex-phase_2024}, the system identification pipeline has been validated on various control designs and experimental setups and has shown consistently good modeling performance. The following section will highlight the results of one specific experiment, including a phase drift. Experimental data were collected from a grid-forming inverter in a power hardware-in-the-loop (PHIL) laboratory (see Fig. \ref{fig:labphoto}). The implemented grid-forming control strategy employs a standard droop scheme\cite{schiffer_conditions_2014}.

\begin{figure}[htbp]
        \centering
        \includegraphics[width=\linewidth]{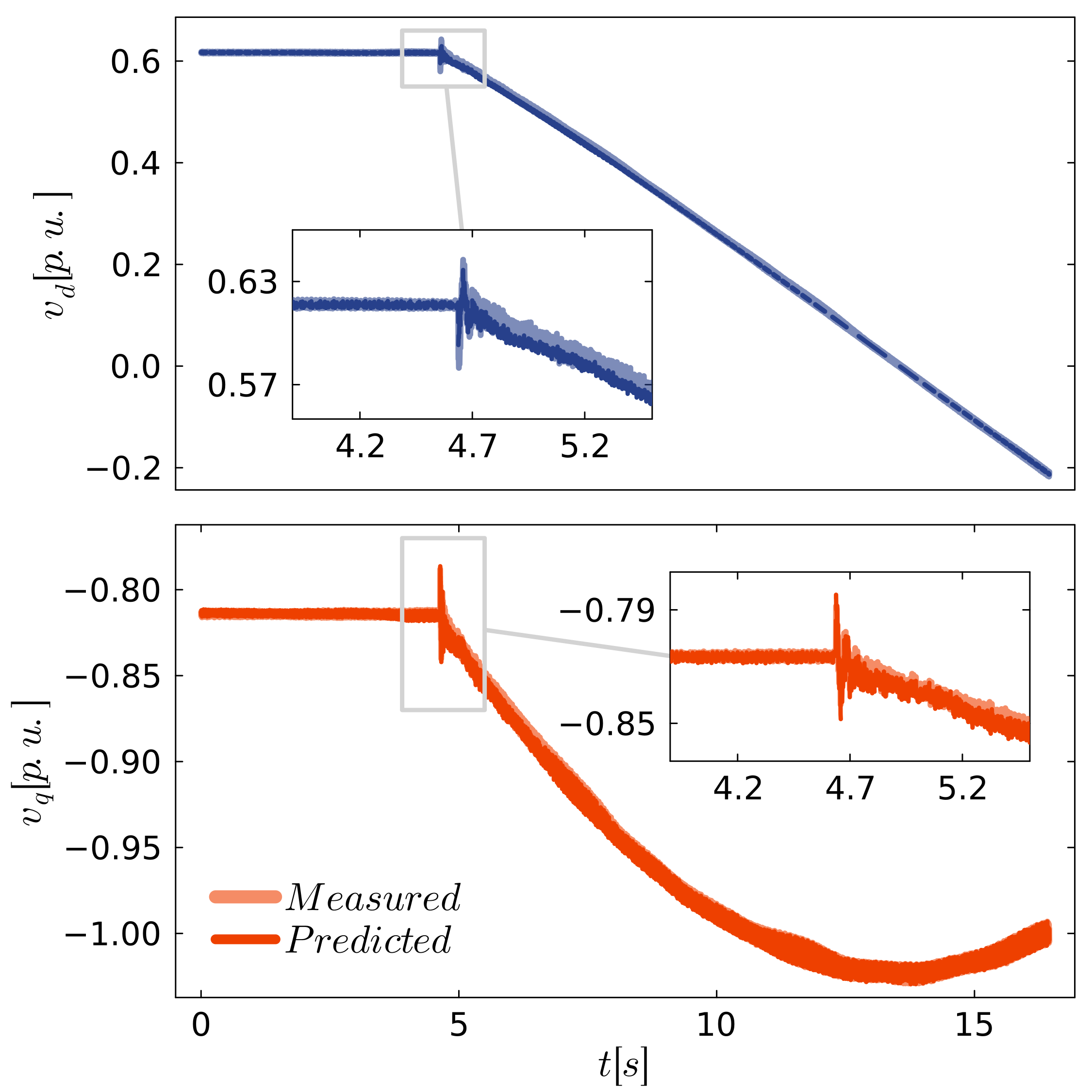}
        \caption{Performance for the measured and predicted $dq$-voltage for the droop-controlled inverter. The lens shows the disconnection event.}
        \label{fig:out_of_distribution_lab}
    \end{figure}

The inverter dynamics were identified by connecting an individual inverter to a slack bus with a time-varying voltage phase and magnitude. We then performed an "out-of-distribution" experiment, confronting the inverters with a situation far from the data used to identify them. This experimental setup consists of two droop-controlled grid-forming inverters and a load configured within a microgrid, initially connected to an auxiliary grid. At a discrete event, the connection to the auxiliary grid is severed, necessitating the grid-forming inverters to compensate for the resulting power mismatch. This power mismatch leads to a phase drift, observable in Fig.~\ref{fig:out_of_distribution_lab} after the disconnection occurs at $t = 4.5$ s. As discussed in section \ref{sec:problem}, such a phase drift cannot be accurately described using a single linearized model in conventional coordinates. The normal form framework captures this behavior by eliminating the dependency on a phase reference.

    Fig.~\ref{fig:out_of_distribution_lab} depicts the predicted and measured $dq$-voltages during the islanding process. Remarkably, the identified normal form with only a single internal variable accurately predicts the $dq$-voltage transients, underscoring its capabilities in modeling grid-forming inverters.

The fact that the identified model performs so well during a rather extreme perturbation such as a sudden line failure and disconnection, is highly promising. For more detailed results on the performance of this approach, and further simulation studies, we refer to \cite{buttner_complex-phase_2024}.

\section{Stability analysis}
\label{sec:stability analysis}

In \cite{niehues_small-signal_2024}, the complex phase has been used to obtain conditions for the small-signal stability of a wide range of power grid models.
The power system can be represented as a feedback loop between grid-forming devices that take $(\delta Q, \delta P)$ as input and provide $(\dot\sigma, \dot \varphi)$ as the output, and transmission lines that provide $(\delta Q, \delta P)$ from $(\dot\sigma, \dot \varphi)$.
The small-signal stability of the interconnected system was analyzed in terms of the transfer matrices of both subsystems using a variant of the small-phase theorem \cite{chen_phase_2024}.

The LTI system given in equations \eqref{eq:normal form Hammerstein Wiener x}-\eqref{eq:normal form Hammerstein Wiener eta} can be cast into transfer matrix form:
\begin{align}
    &T(s) := \bmat{T_{\rho P}(s) & T_{\rho Q}(s) & T_{\rho V}(s)\\ T_{\omega P}(s) & T_{\omega Q}(s) & T_{\omega V}(s)},
\end{align}
where $s$ is the Laplace frequency. The voltage magnitude depends non-linearly on the real part of the complex phase $\sigma$, but when linearizing, we see that $\delta V = V^\circ \delta \sigma$. Hence, if there are no internal states and thus $T$ does not depend on $s$, $T$ recovers the form \eqref{eq:linearized complex frequency dynamics invariance}.

For our analysis, we assume that the system implements a commonly used droop between $V$ and $Q$, such that:
\begin{align}
    \bmat{T_{\rho Q}(s)\\ T_{\omega Q}(s)} = \tilde\alpha V^\circ     \bmat{T_{\rho V}(s)\\ T_{\omega V}(s)}
\end{align}
for some $\tilde\alpha \in\mathbb{R}$.
Now, \eqref{eq:normal form Hammerstein Wiener x}-\eqref{eq:normal form Hammerstein Wiener eta} can be cast into a convenient form:
\begin{align}\label{eq:nodal transfer matrix}
    s \bmat{\delta\sigma\\ \delta \varphi}
    = \bmat{\delta \rho\\ \delta \omega}
    = \bmat{T_{\rho Q} & T_{\rho P} \\ T_{\omega Q} & T_{\omega P}} \bmat{\delta Q + \tilde\alpha \delta\sigma \\ \delta P}.
\end{align}
As noted above, the complex phase formulation means that these transfer matrices do not depend on an arbitrary phase or reference frame. They only depend on the power flow and the device dynamics, and directly encode how relative amplitude velocity $\rho$ and frequency $\omega$ react to active and reactive power imbalances $\delta P$, $\delta Q$.

Using these advantages, in \cite{niehues_small-signal_2024} the authors have shown that an operating point in a system of  heterogeneous devices, all with transfer matrices of the form \eqref{eq:nodal transfer matrix} and connected by inductive transmission lines, is stable under the following conditions:
Consider a lossless power grid with admittance Laplacian $\bm Y$ such that the current is given by $\bm \imath = \bm Y \bm v$, where $\bm\imath$ and $\bm v$ have components $\hat\imath_n$ and $\hat v_n$ for bus $n$, respectively. Let the operating point be a power flow solution with voltage phase angles ${\varphi}^\circ_n$ and magnitudes $V_n^\circ = e^{{\sigma_n}\!\!\!^\circ}$. To specify static operational bounds, denote the maximum voltage ratio \textnormal{$\gamma_\text{max}$} and maximum phase difference \textnormal{$\Delta\varphi_\text{max}$} such that \textnormal{$\gamma^{-1}_\text{max} < V_n^\circ/V_m^\circ < \gamma_\text{max}$} and \textnormal{$|\varphi_n^\circ - \varphi_m^\circ| < \Delta\varphi_\text{max} < \pi /2$} for all $n$ and $m$ connected by a line.

If the small-signal response of all buses can be described by \eqref{eq:nodal transfer matrix} (with $T_\bullet^n$ for bus $n$), then the operating state is linearly stable if for all buses $n$ and Laplace frequencies $s$:
    \begin{align}
        \Re(T^n_{\rho Q}(s)) < 0 \; , &\quad \Re(T^n_{\omega P}(s))  < 0\, ,
        \label{eq:tr_T_nodes_definite}
        \\
        \Re(T^n_{\rho Q}(s)) \cdot \Re(T^n_{\omega P}(s)) &< \frac{1}{4}\left|T^n_{\rho P}(s) +  T^{n*}_{\omega Q}(s)\right|^2\, ,
        \label{eq:det_T_nodes_positive}
        \\
        \tilde\alpha_n \ge 2 {V_n^\circ}^2 |Y_{nn}| &\left( \frac{\gamma_\text{\textnormal{max}}}{\cos\Delta\varphi_\text{\textnormal{max}}} -1 \right)\, .\label{eq:alpha_bound_T_lines_definite}
    \end{align}

These conditions are straightforward to interpret. The first one requires that the frequency reaction $\delta\omega$ to active power mismatch $\delta P$ and the voltage reaction $\delta \rho$ to reactive power mismatch $\delta Q$ point in the right direction, i.e., they have negative signs.
The second line requires that this stabilizing reaction of frequency to power, and amplitude to reactive power mismatch dominates the cross talk, that is, the reaction of amplitude to active power and frequency to reactive power.
The final condition requires that the coefficient in the $V$-$Q$ droop law, which governs the intrinsic voltage stability, is sufficiently large.
These conditions correspond well with general design principles for grid-forming inverters. If the transfer functions are just coefficients independent of $s$, similar conditions have long been known. The above conditions thus represent a vast generalization of these established conditions to devices with very general internal structure.
For established models, the transfer functions can be easily related to machine parameters.
Thus, \eqref{eq:tr_T_nodes_definite}-\eqref{eq:alpha_bound_T_lines_definite} were used to reproduce known stability conditions and gain new insights through new stability conditions \cite{niehues_small-signal_2024}.

\section{Discussion}
\label{sec:discussion}
In this paper, we have shown that the linearization of power grid dynamics around time-varying operating states results in a linear-time-independent system if complex phase coordinates are used.
These are much easier to analyze than the linear-time-periodic systems acquired in conventional $\alpha\beta$-coordinates, and do not depend on a moving reference frame as $dq$-coordinates do.
Moreover, the complex phase linearization is independent of the reference phase given by the linearization point, and linearized phase shifts and non-linear phase shifts coincide.
This makes complex phase coordinates especially robust, as the linear system stays valid during phase drifts often experienced in real-world situations.
We have shown that these advantages enable successful system identification, in which general dynamical characteristics can be extracted from data.
These characteristics can also be used for the general stability analysis independent of reference points.

In conclusion, the complex phase is a promising tool for modeling and analyzing grid-forming inverters in heterogeneous grids and understanding their collective behavior.

For future work it would be interesting to perform an in-depth comparison of system identification as in \cite{buttner_complex-phase_2024} in different variables, for a variety of systems and settings. Regarding the small-signal stability analysis in \cite{niehues_small-signal_2024}, it is feasible to make these results robust in the $\mathcal{H}_\infty$ sense, and an extension to grids with homogeneous losses is straightforward. However, it remains an open question how to extend the results to more general models that don't implement an exact $V$-$Q$ droop, or to settings with inhomogeneous losses.







\end{document}